# Cactus Graphs and Some Algorithms

*Kalyani Das*
Ramnagar College, Depal, Purba Medinipur
Midnapore – 721102, India.
email: kalyanid380@gmail.com

**Abstract.** A cactus graph is a connected graph in which every block is either an edge or a cycle. In this paper, we consider several problems of graph theory and developed optimal algorithms to solve such problems on cactus graphs. The running time of these algorithms is $O(n)$, where n is the total number of vertices of the graph. The cactus graph has many applications in real life problems, especially in radio communication system.

## 1. Introduction

In graph theory, a cactus graph is a connected graph in which any two simple cycles have at most one vertex in common. Equivalently, every edge in such a graph belongs to at most one simple cycle. Equivalently, every block (maximal subgraph without a cutvertex) is an edge or cycle.

Let $G = (V, E)$ be a finite, connected, undirected simple graph of $n$ vertices $m$ edges, where $V$ is the set of vertices and $E$ is the set of edges. A vertex $u$ is called a *cutvertex* if removal of $u$ and all edges incident on $u$ disconnect the graph. A connected graph without a cutvertex is called a *non-separable* graph. A *block* of a graph is a maximal non-separable subgraph. A *cycle* is a connected graph (or subgraph) in which every vertex is of degree two. A block which is a cycle is called a *cycliced block*. A *cactus graph* is a connected graph in which every block is either an edge or a cycle. A *weighted* graph $G$ is a graph in which every edge is associates with a weight. Without loss of generality we assume that all weights are positive. A *weighted cactus graph* is a weighted, connected graph in which every block containing two vertices is an edge and three or more vertices is a cycle.

Cactus graph were first studied under the name of Husimi trees, bestowed on them by Frank Harary and George Eugene Unlenbeck in honour of previous work of these graphs by Kodi Husimi. Cactus graph has many applications. These graphs can be used to model physical setting where a tree would be inappropriate. Examples of such setting arise in telecommunications when considering feeder for rural, suburban and light urban regions [33] and in material handling network when automated guided vehicles are used [34]. Moreover, ring and bus structures are often used in local area networks. The combination of local area network forms a cactus graph.

Because of various applications in real life situation and telecommunication problem, cactus graphs have extensive studied during last decade. Some well known problems like all-pair shortest path problem, domination problem, coloring and labeling problems, covering problems etc. are solved in polynomial time on cactus graphs efficiently. Lot of algorithms have been design to solve various graph theoretic problems, some of them are available in [48-59].

To solve some problems on cactus graphs, a tree is constructed, called $T_{BC}$ tree, which is described below.

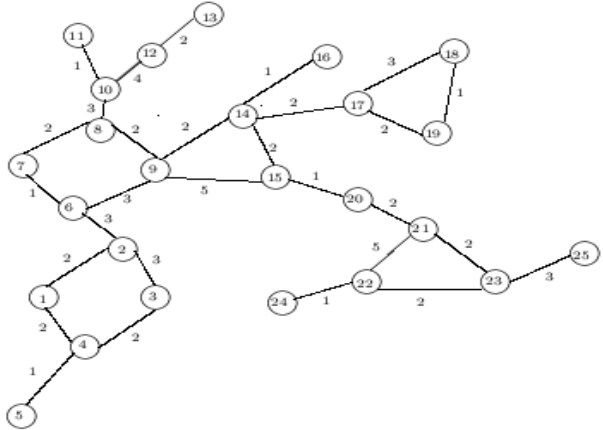

**Figure 1:** A weighted cactus graph $G$.

## 2. Formation of the tree $T_{BC}$

In this thesis we use a method in which blocks and cutvertices of the graph $G$ are determined using DFS technique and there after form an intermediate graph $G'$ i.e., $G' = (V', E')$ where $V' = \{B_1, B_2, \ldots, B_N\}$ and $E' = \{(B_i, B_j) : i \neq j, \quad i, j = 1, 2, \ldots, N, \quad B_i$ and $B_j$ are adjacent blocks $\}$.

Now the tree $T_{BC}$ is constructed from $G'$ as follows:

We discard some suitable edges from $G'$ in such a way that the resultant graph becomes a tree. The procedure for such reduction is given below:

Let us take any arbitrary vertex of $G'$, containing at least two cutvertices of $G$, as root of the tree $T_{BC}$ and mark it. All the adjacent vertices of this root are taken as children of level one and are marked. If there are edges between the vertices of this level, then those edges are discarded. Each vertices of level one is considered one by one to find the vertices which are adjacent to them but unmarked. These vertices are taken as children of the corresponding vertices of level one and are placed at level two. These children at level two are marked and if there be any edge between them then they are discarded. This process is continued until all the vertices are marked.

Thus the tree $T_{BC} = (V', E'')$ where $V' = \{B_1, B_2, \ldots, B_N\}$ and $E'' \subset E'$ is obtained.

For convenience, we refer the vertices of $T_{BC}$ as nodes. We note that each node of this tree is a block of the graph $G = (V, E)$. The parent of the node $B_i$ in the tree $T_{BC}$ will be denoted by *Parent($B_i$)*.

## 3. Different Problems on Cactus Graphs and its Solutions
### 3.1. Computation of All-Pairs Shortest Paths on Weighted Cactus Graphs

Let $G = (V, E)$ be a finite, connected, undirected, simple graph of $n$ vertices and $m$ edges, where $V$ is the set of vertices and $E$ is the set of edges. A *path* of a graph $G$ is an alternating sequence of distinct vertices and edges which begins and ends with vertices in $G$. The *length* of a path is the sum of the weights of the edges in the path. A path from vertex $u$ to $v$ is a *shortest path* if there is no other path from $u$ to $v$ with lower length. The *distance* $d(u,v)$ between vertices $u$ and $v$ is the length of shortest path between $u$ and $v$ in $G$.

For any general graph with $n$ vertices, solution to the all-pair shortest path problem takes $O(n^3)$ time [1]. A lot of work have been done in improving this running time using randomization and probabilistic methods for general as well as special kinds of graphs. Ahuja et.al. [2] have given a faster sequential using Radix heap and Fibonacci heap for single source shortest path problem in $O(m+n\sqrt{logC})$ time for a network with $n$ vertices and $m$ edges and non-negative integers are costs bounded by $C$. In [47], Seidal has given an $O(M(n)logn)$ time sequential algorithm for all-pair shortest path problem for an undirected and unweighted arbitrary graph with $n$ vertices, where $M(n)$ is the time (best value of $M(n)$ is $O(n^{2.376})$ necessary to multiply two $n \times n$ matrices of small integers.

Alon et. al. [3] have reported a sub-cubic algorithm for computing APSP on directed graph with edge length which require $O(Mn^{\gamma})$ time, where $\gamma = (3+\omega)/2$, $\omega < 3$ and $M$ is the largest edge length. Galil and Margalit [16] have improved the dependence of $M$ and have also given an $O(M^{(\omega+1)/2}n^{\omega}logn)$ algorithm for undirected graph. Ravi et. al. [42] have given a sequential algorithm to solve all-pair shortest path (APSP) on interval graph in $O(n^2)$ time. Pal and Bhattacharjee in [41] have given an $O(n^2)$ time algorithm for finding the distance between all pair of vertices on interval graphs.

In this problem, we select a specified vertex $x$ and find a block which contains $x$. We construct a tree $T_{BC}$ taking this block as root. After constructing the tree we first compute the distance between $x$ and all other vertices in this root. Then we compute the distance from $x$ to vertices (other than $x$) of the blocks corresponding to the nodes in level one as follows. Let $B_i$ be a node at level one. Then $e_i$ is its entry point. We compute the distances of every vertex $v$ of $B_i$ from $e_i$ and adding $d(x,e_i)$ with these distances we obtain the distance and so shortest path of the vertices of the block $B_i$ from $x$. Similarly, we compute the distance from $x$ to other vertices of the blocks at level one. For the nodes, *i.e.,* blocks of the remaining levels, the distance from $x$ can be computed by the same process.

In general, let us consider a block $B_j$ at level '$i$' and assume that the distance between $x$ and all vertices of the blocks at level '$i-1$' have been calculated. The entry point of $B_j$ is $e_j$. Then $d(x, e_j)$ is known as $e_j$ belongs to a block at level '$i-1$'. We now compute $d(e_j, v)$ for all $v \in B_j$. Then $d(x, u) = d(x, e_j) + d(e_j, u)$ for all $u \in B_j$. This determine the distance between $x$ and any vertex of any block at level '$i$'.

This procedure takes $O(n)$ time. Thus to compute all-pair shortest path on weighted cactus graphs, $O(n^2)$ time is required.

### 3.2. Finding a Minimum Dominating Set

Let $G = (V, E)$ be a finite, connected, undirected, simple graph of $n$ vertices and $m$ edges, where $V$ is the set of vertices and $E$ is the set of edges. A subset $D$ of $V$ is said to *dominate* $V$ if every vertex in $V - D$ is adjacent to at least one vertex in $D$. In this case $D$ is called a *dominating set* of the graph $G$. The set $D$ is called a *minimum dominating set* if the cardinality of $D$ is minimum among all dominating sets of the graph $G$.

The problem of determining a minimum cardinality dominating set has been discussed in [11], and has obvious application in the optimum location of facilities in a network. When restricted to interval graphs, the minimum dominating set problem along with several related variants, becomes polynomial time solvable [6, 7]. Kratsch et al. [32] first presented polynomial time algorithm which takes $O(n^6)$ time for domination problems on cocomparability graphs. These algorithms are valid for the cardinality case only. In [43], we get a fast algorithm for domination problems on permutation graphs which takes $O(m+n)$ time.

In this problem we construct a tree $T_{BC}$ using the blocks and cutvertices and applying the Euler Tour we obtain a sequence of nodes. Consider the nodes of that sequence one by one and find the dominating vertices from each node.

Hence we form an algorithm which describes a traversing from leaf node to the root. If there exist any subtree on the way of traversing then we traverse all the branches of the subtree from leaf to root except its root and meet the root when all the branches are traversed. For any leaf node and for an interior node we apply different method and obtain the dominating sets for each node.

This process takes $O(n)$ time for computing minimum dominating set on cactus graph $G$.

### 3.3. Finding a Minimum 2-Neighbourhood Covering set

The k-neighbourhood-covering ($k$-NC) problem is a variant of the domination problem. A vertex $x$ $k$-dominates another vertex $y$ if $d(x, y) \leq k$. A vertex $z$ is $k$-NC of an edge $(x, y)$ if $d(x, z) \leq k$ and $d(y, z) \leq k$ *i.e*, the vertex $z$ $k$

Some Algorithms on Cactus Graphs

-dominates both $x$ and $y$. Conversely if $d(x,z) \leq k$ and $d(y,z) \leq k$ then the edge $(x, y)$ is said to be $k$-neighbourhood covered by the vertex $z$. A set of vertices $C \subseteq V$ is a $k$-NC set if every edge in $E$ is $k$-NC by some vertices in $C$. The $k$-NC number $\rho(G,k)$ of $G$ is the minimum cardinality of all $k$-NC sets.

For $k = 1$, Lehel et al. [35] have presented a linear time algorithm for computing $\rho(G,1)$ for an interval graph $G$. Chang et al. [12] and Hwang et al. [22], have presented linear time algorithms for computing $\rho(G,1)$ for a strongly chordal graph provided that strong elimination ordering is known. Hwang et al. [22] also proved that $k - NC$ problem is NP-complete for chordal graphs. Mondal et al. [38] have presented a linear time algorithm for computing 2-NC problem for an interval graph. Also a linear time algorithm for trapezoid graph has presented by Ghosh et al. [17].

In this Problem we construct a tree $T_{BC}$ using blocks and cutvertices of $G$. Thereafter applying Euler tour on that tree we obtain a sequence of nodes. There are two types of nodes, some are leaf nodes and some are interior nodes. Depending upon the number of vertices of cycles and paths we determine the number of covering vertices from each node as well as the graph $G$.

Thus the algorithm which finds the 2-neighbourhood covering (2-NC) set of the graph $G$ in $O(n)$ time. The algorithm also takes $O(n)$ space.

### 3.4. Finding a Maximum Weight 2-Colour Set on Weighted Cactus Graphs

The graph colouring problem (GCP) plays a central role in graph theory and it has direct applications in real life problems [5], and is related to many other problems such as timetabling [13, 37], frequency assignment [19] etc. A k-colouring (assignment) of an undirected graph $G = (V, E)$, where V is the set of $|V| = n$ vertices and $E \subseteq V \times V$ the set of edges, is a mapping $\alpha : V \to \{1, 2, \ldots, k\}$ that assigns a positive integer from $\{1, 2, \ldots, k\}$ (representing the colours) to each vertex. We say that a colouring is feasible if the end nodes of every edge in E have assigned different colours, *i.e.*, for all $(u, v) \in E$, $\alpha(u) \neq \alpha(v)$. We call conflict the situation when two nodes between which an edge exists have the same colour associated to them. We say that a colouring is infeasible if at least one conflict occurs. Alternatively to the formulation as an assignment problem, the GCP can also be represented as a partitioning problem, in which a feasible k-colouring corresponds to a partition of the set of nodes into k sets $C_1, C_2, \ldots, C_k$ such that no edge exists between two nodes from the same colour class.

The graph colouring problem is NP-complete. Hence, we need to use approximate algorithmic methods to obtain solutions close to the absolute minimum in a reasonable execution time.

The maximum weight k-colourable Subgraph (MWKC) problem is related to the following problem. The input to this problem consist of an integer number $k$ and an undirected graph $G = (V, E)$, where each vertex $v$ has a non-negative weight $w_v$.

The goal is to pick a subset $V' \subseteq V$, such that there exists a colouring $c$ of $G[V']$ with $k$ colours, and among all such subsets, the value $\sum_{v \in V'} w_v$, $w_v$ is maximum. This problem is NP-hard for general graph even for split graph [20].

The maximum weight k-colouring problem is same as the maximum weight k-independent set (MWKIS) problem. The maximum k-independent set problem on $G$ is to determine $k$ disjoint independent sets $S_1, S_2, \ldots, S_k$ in $G$ such that $S_1 \bigcup S_2 \bigcup \ldots S_k$ is maximum. The MWKIS problem is NP-complete for general graphs [20].

Many work on colouring problem has been done previously. Local search in large neighbour and iterated local search for GCP are described in [9, 4]. The maximum weight 2-colouring problem or the maximum weight 2-independent set (MW2IS) problem, which is a special case of the (MWKIS) problem, is also NP-complete for general graphs and it applications have been studied in the last decade [23, 24, 36]. In [23], Hsiao et.al. have solved the two-track assignment problem by solving the M2IS problem on circular arc graph. In [36], Lou et. al. have solved the maximum 2-chain problem on a given point set, which is the same as the MW2IS problem on permutation graph.

In this problem we find odd and even blocks and form block-cutvertex graph $G''$ using the odd blocks only from the graph G. Next represent $G''$ in terms of edge weight(weight of the cutvertex) and vertex weight(weight of the minimum weight vertex) and form a tree $T_{BQ}$. The method of finding maximum weight 2-coloured set is to delete such a vertex from each odd block so that minimum weight is dicarded. Also in this problem we find minimum weight feedback vertex set. Here we select minimum weight vertices or cutvertex from both even and odd blocks.

Thus the algorithm which finds the 2-coloured set as well as the minimum feedback vertex of the graph $G$ takes $O(n)$ time. The algorithm also takes $O(n)$ space.

### 3.5. Finding a Maximum Independent Set and Maximum 2-Independent Set

Let $G = (V, E)$ be a finite, connected, undirected, simple graph of $n$ vertices and $m$ edges, where $V$ is the set of vertices and $E$ is the set of edges. A subset of the vertices of a graph $G = (V, E)$ is an independent set if no two vertices in this subset are adjacent. The maximum independent set (MIS) problem on $G$ is to determine a maximum size independent set on $G$. The MIS problem is NP-complete for general graphs [18], but it can be solved in polynomial time for many special graphs [28].

The maximum k-independent set (MKIS) problem on $G$ is to determine $k$ disjoint independent sets $S_1, S_2, \ldots, S_k$ in $G$ such that $S_1 \bigcup S_2 \bigcup \ldots S_k$ is maximum. The MKIS problem is NP-complete for general graphs [20].

The maximum 2-independent set (M2IS) problem, which is a special case of the MKIS problem, is also NP-complete for general graphs and it applications have been studied in the last decade [23, 36]. In [23], Hsiao et. al. have solved the two-track assignment problem by solving the M2IS problem on circular arc graph. In [36], Lou et. al. have solved the maximum 2-chain problem on a given point set, which is the same as the M2IS problem on permutation graph.

In this problem, MKIS problem is considered on a non-weighted cactus graph for $k = 1$ and $k = 2$.

In this problem a tree $T_{BC}$ is constructed using blocks and cutvertices of the graph $G$. There after apply Euler Tour to find a sequence of nodes to consider one by one from leaf to root node. For leaf nodes and interior nodes separate techniques are used to find vertices for independent set. For the 2-independent set problem one vertex from each odd cycle is removed so that alternate vertices from cycles and paths of the graph $G$ form 2-independent set.

Thus the algorithms for the above two problems take $O(n)$ time.

### 3.6. Finding Maximum and Minimum Height Spanning Trees

Let $G = (V, E)$ be a finite, connected, undirected, simple graph of $n$ vertices and $m$ edges, where $V$ is the set of vertices and $E$ is the set of edges. A tree is a connected graph without any circuits. A tree $T$ is said to be a spanning tree of a connected graph $G$ if $T$ is a subgraph of $G$ and $T$ contains all vertices of $G$. The *longest distance* $ld(u,v)$ and *distance* $d(u,v)$ between two vertices $u$ and $v$ are the length $lp(u,v)$ and $\rho(u,v)$ in $G$ if such paths exist.

Note that $ld(u,u) = 0$, $ld(u,v) = ld(v,u)$ and $ld(u,v) \leq ld(u,w) + ld(w,v)$. Also $d(u,u) = 0$, $d(u,v) = d(v,u)$ and $d(u,v) \leq d(u,w) + d(w,v)$.

The *elongation* of a vertex $u$ in a graph $G$ is the longest distance from vertex $u$ to a vertex furthest from $u$ i.e., $el(u) = max\{ld(u,v) : v \in V\}$. Vertex $v$ is said to be a *furthest vertex* of $u$ if $ld(u,v) = el(u)$.

The *eccentricity* of a vertex $u$ in a graph $G$ is the longest distance from the vertex $u$ to a vertex furthest from $u$ i.e, $e(u) = max\{d(u,v) : v \in V\}$.

In a tree, a vertex $v$ is said to be at *level* $l$ if $v$ is at a distance $l$ from the root. The *height* of a tree is the maximum level which is occurred in the tree.

A graph may have more than one spanning tree. The height of a spanning tree $T$ of a graph $G$ is denoted by $H(T, G)$. A *maximum height spanning tree* is a spanning tree whose height is maximum among all spanning trees of a graph. The height of the maximum height spanning tree of a graph $G$ is denoted by $H_{max}(G) = max\{el(u) : u \in V\}$.

Suppose $v$ be the vertex for which $H_{max}(G)$ is attained and $v'$ its furthest vertex, then the longest path *i.e.*, $lp(v, v')$ is called as *maximum height path* $(v, v')$ and denoted by $MHP(v, v')$.

A *minimum height spanning tree* is a spanning tree whose height is minimum among all spanning tree of a graph. The height of the minimum height spanning tree of a graph $G$ is denoted by $H_{min}(G) = min\{e(u) : u \in V\}$. The vertex $x$ for which $H_{min}(G) = e(x)$ is called the *center* of $G$.

Some related works are discussed here: In [44], a spanning tree of maximal weight and bounded radius is determined from a complete non-oriented graph $G = (V, E)$ with vertex set $V$ and edge set $E$ with edge weight in $O(n^2)$ time, $n$ is the total number of vertices in $G$. In [39], the minimum spanning tree problem is considered for a graph with $n$ vertices and $m$ edges. They introduced randomized search heuristics to find minimum spanning tree in polynomial time with out employing global techniques of greedy algorithms. In [26], the authors find a spanning tree $T$ that minimizes $D_T = Max_{(i,j) \in E} d_T(i, j)$ where $d_T(i, j)$ is the distance between $i$ and $j$ in a graph $G = (V, E)$. The minimum restricted diameter spanning tree problem is to find spanning tree $T$ such that the restricted diameter is minimized. It is solved in $O(\log n)$ time. In [27], the minimum diameter spanning tree problem on graphs with non-negative edge lengths is determined which is equivalent for finding shortest paths tree from absolute 1-center problem of the general graph is solvable in $O(mn + n^2 \log n)$ time [34].

In this Problem, we find the maximum height spanning tree by finding the elongation and the longest path $MHP(u, v)$. Then deleting one edge from each cycle which is not consider during the calculation of elongation and $MHP(u, v)$ the maximum height of the spanning tree is obtained whose height is equal to $MHP(u, v)$. Also we find the minimum height spanning tree by finding the eccentricity and the radius of the graph $G$ and deleting one edge from each cycle so that the radius is the minimum height of the spanning tree.

These algorithms find the maximum height spanning tree and minimum height spanning tree in $O(n)$ time.

### 3.7. L(2,1)-labelling of cactus graphs

The $L(2,1)$-labelling of a graph $G$ is an abstraction of assigning integer frequencies to radio transmitters such that the transmitters that are one unit of distance apart receive frequencies that differ by at least two, and transmitters that are two units of distance apart receive frequencies that differ by at least one. The span of an $L(2,1)$-labelling is the difference between the largest and the smallest frequencies assigned to the vertices. The $L(2,1)$-labelling number of a graph $G$, denoted by $\lambda(G)$, is the least integer $k$ such that $G$ has an $L(2,1)$-labelling of span $k$.

Several results are known for $L(2,1)$-labelling of graphs, but, to the best of our knowledge no result is known for cactus graph.

The lower bound for $\lambda(G)$ is $\Delta + 1$, which is achieved for the star $K_{1,\Delta}$. Griggs and Yeh [15] prove that $\lambda(G) \leq \Delta^2 + 2\Delta$ for general graph and improve this upper

Some Algorithms on Cactus Graphs

bound to $\lambda(G) \leq \Delta^2 + 2\Delta - 3$ when $G$ is 3- connected and $\lambda(G) \leq \Delta^2$ when $G$ is diameter 2 (diameter 2 graph is a graph where all nodes have either distance 1 or 2 each other). Jonas [29] improves the upper bound to $\lambda(G) \leq \Delta^2 + 2\Delta - 4$ if $\Delta \geq 2$, by constructive labelling schemes. Chang and Kuo [10] further decrease the bound to $\Delta^2 + \Delta$. Further, Kral and Skrekovski [31] improves this bound $\lambda(G) \leq \Delta^2 + \Delta - 1$ for any graph $G$. The best known result till date is $\lambda(G) \leq \Delta^2 + \Delta - 2$ due to Goncalves [14]

To label the vertices of a cactus graph, we first label the vertices of all induced subgraphs of the cactus graph. We obtained the following results.

Let $H$ be a subgraph of $G$, then obviously $\lambda(H) \leq \lambda(G)$ [10].

If $G$ and $H$ are two graphs and if $V_G \cap V_H = \phi$ then
$$\lambda(G \bigcup H) = max\{\lambda(G), \lambda(H)\} \text{ and}$$
$$\lambda(G + H) = max\{|V_H| - 1, \lambda(G)\} + max\{|V_H| - 1, \lambda(H)\} + 2 \text{ [10]}.$$
Also,
$$\lambda(G \bigcup_v H) \geq max\{\lambda(G), \lambda(H)\}, \text{ where } \{v\} = V_G \cap V_H.$$

For any star graph $K_{1,\Delta}$, $\lambda(K_{1,\Delta}) = \Delta + 1$, which is equal to $n$, where $n$ is the number of vertices.

For any cycle $C_n$ of length $n$, $\lambda(C_n) = 4 = \Delta + 2$ [15].

Suppose a graph G contains two cycles $C_n$ and $C_m$ joined by a cutvertex $v_0$, then $\lambda(C_n \bigcup_{v_0} C_m) = 5 = \Delta + 1$.

Let a graph $G_1$ contains $n$ number of triangles with a common cutvertex. Then $\lambda(G_1) = \Delta + 1$ or $\Delta + 2$ according as $n$ is even or odd, where $\Delta$ is the degree of the cutvertex.

Let a graph $G$ contains $n$ number of cycles of length 3 and $m$ number of cycles of length 4. If they have a common cutvertex with degree $\Delta$, then $\lambda(G) = \Delta + 1$.

Let $G$ be a graph which contains finite number of cycles of any length and finite number of edges. If $v_0$ be the common cutvertex with degree $\Delta$ then $\lambda(G) = \Delta + 1$.

Let $G$ be a graph, contains a cycle of any length and finite number of edges, they have a common cutvertex $v_0$. If $\Delta$ be the degree of the cutvertex then, $\lambda(G) = \Delta + 1$.

Let $G$ be a graph contains a cycle of any length and each vertex of the cycle has another cycle of length three. If $\Delta$ is the degree of $G$ then $\lambda(G) = \Delta + 3$.

For any caterpillar graph the value of $\lambda$ lies between $\Delta + 1$ and $\Delta + 2$.

Let $G_1$ and $G_2$ be two cactus graphs. If $\Delta_1 + 1 \leq \lambda(G_1) \leq \Delta_1 + 3$ and

$\Delta_2 + 1 \leq \lambda(G_2) \leq \Delta_2 + 3$, then, $\Delta + 1 \leq \lambda(G) \leq \Delta + 3$, where $G = G_1 \bigcup_v G_2$.

The time complexity of the proposed algorithm to label a cactus graph using L(2,1)-labelling technique takes $O(n)$ time, where n is the total number of vertices of the cactus graph.

### 3.8. L(0,1)-labelling of cactus graphs

An $L(0,1)$-labelling of a graph $G$ is an assignment of nonnegative integers to the vertices of $G$ such that the difference between the labels assigned to any two adjacent vertices is at least zero and the difference between the labels assigned to any two vertices which are at distance two is at least one. The span of an $L(0,1)$-labelling is the maximum label number assigned to any vertex of $G$. The $L(0,1)$-labelling number of a graph $G$, denoted by $\lambda_{0,1}(G)$, is the least integer $k$ such that $G$ has an $L(0,1)$-labelling of span $k$. This labelling has an application to a computer code assignment problem. The task is to assign integer control codes to a network of computer stations with distance restrictions.

Some results are available on $L(h,k)$-labelling problem. Here we discuss some particular cases. When $h = 0$ and $k = 1$ then we get $L(0,1)$-labelling problem. Several results are known for $L(0,1)$-labelling of graphs, but, to the best of our knowledge no result is known for cactus graph. In this section, the known result for general graphs and some related graphs of cactus graph are presented.

The upper bound for $\lambda_{0,1}(G)$ of any graph $G$ is $\lambda_{0,1}(G) \leq \Delta^2 - \Delta$ [30], where $\Delta$ is the degree of the graph.

Here we label the vertices of a cactus graph by $L(0,1)$-labelling and have shown that, $\Delta - 1 \leq \lambda_{0,1}(G) \leq \Delta$ for a cactus graph, where $\Delta$ is the degree of the graph $G$. Here we start the labelling by the labelling the subgraphs of the cactus graph. And we obtained some results which are stated below.

If we label a star graph $K_{1,\Delta}$ by $L(0,1)$-labelling, then we get $\lambda_{0,1}(K_{1,\Delta}) = \Delta - 1$. For any cycle $C_n$ of length $n$, $\lambda_{0,1}(C_n) = 1$, when $n = 4k$, where $k$ is a positive integer, and $\lambda_{0,1}(C_n) = 2$ for other cases [8].

Let $G$ be a graph which contains two cycles and they have a common cutvertex. If $\Delta$ be the degree of G, then, $\lambda_{0,1}(G) = \Delta$, when two cycles are of length 3 and $\Delta - 1$, for others. This result is true for the graph contains $n$ number of cycles of any lengths, joined with a common cutvertex.

For the graph which contains finite number of cycles of any length and finite number of edges, then $\lambda_{0,1}(G) = \Delta - 1$. If the graph is a sun graph with $2n$ vertices, then we proved that $\lambda_{0,1}(S_{2n}) = 2 = \Delta - 1$. Suppose $G$ contains a cycle of any length and each vertex of the cycle has another cycle of any length, then

$\Delta - 1 \leq \lambda_{0,1}(G) \leq \Delta$.

It is proved for caterpillar, lobster and tree the value of $\lambda_{0,1}$ is $\Delta - 1$.

Finally, by arranging all the results, we can conclude that for a cactus graph $\Delta - 1 \leq \lambda_{0,1}(G) \leq \Delta$.

### 3.9. $(2,1)$-total labelling of the cactus graph

A $(2,1)$-total labelling of a graph $G = (V, E)$ is an assignment of integers to each vertex and edge such that: (i) any two adjacent vertices of $G$ receive distinct integers, (ii) any two adjacent edges of $G$ receive distinct integers, and (iii) a vertex and its incident edge receive integers that differ by at least 2. The *span* of a $(2,1)$-total labelling is the maximum difference between two labels. The minimum span of a $(2,1)$-total labelling of $G$ is called the $(2,1)$-total number and denoted by $\lambda_2^t(G)$.

Motivated by frequency channel assignment problem Griggs and Yeh [15] introduced the $L(2,1)$-labelling of graphs. The notation was subsequently generalized to the $L(p,q)$-labelling problem of graphs. Let $p$ and $q$ be two non-negative integers. An $L(p,q)$-labelling of a graph $G$ is a function $c$ from its vertex set $V(G)$ to the set $\{0,1,\ldots,k\}$ such that $|c(x)-c(y)| \geq p$ if $x$ and $y$ are adjacent and $|c(x)-c(y)| \geq q$ if $x$ and $y$ are at distance 2. The $L(p,q)$-labelling number $\lambda_{p,q}(G)$ of $G$ is the smallest $k$ such that $G$ has an $L(p,q)$-labelling $c$ with $\max\{c(v) | v \in V(G)\} = k$.

This labelling is called $(2,1)$-total labelling of graphs which introduced by Havet and Yu [21] and generalized to the $(d,1)$-total labelling, where $d \geq 1$ be an integer. A $k$-$(d,1)$-total labelling of a graph $G$ is a function $c$ from $V(G) \cup E(G)$ to the set $\{0,1,\ldots,k\}$ such that $c(u) \neq c(v)$ if $u$ and $v$ are adjacent and $|c(u)-c(e)| \geq d$ if a vertex $u$ is incident to an edge $e$. The $(d,1)$-total number, denoted by $\lambda_d^t(G)$, is the least integer $k$ such that $G$ has a $k$-$(d,1)$-total labelling.

It is shown in [40] that for any cactus graphs, $\Delta + 1 \leq \lambda_{2,1} \leq \Delta + 3$. Now in this section, we label the vertices and edges of a cactus graphs $G$ by $(2,1)$-total labelling and it is shown that $\Delta + 1 \leq \lambda_2^t \leq \Delta + 2$ [21].

We label the vertices and edges of a cactus graph by $(2,1)$-total labelling procedure and have shown that, $\Delta + 1 \leq \lambda_2^t(G) \leq \Delta + 2$ for a cactus graph, where $\Delta$ is the degree of the graph $G$. First we label the vertices of different subgraphs of cactus graph by $(2,1)$-total labelling.

If $H$ is a subgraph of $G$, then $\lambda_2^t(H) \leq \lambda_2^t(G)$. For any star graph $K_{1,\Delta}$,

$\lambda_2^t(K_{1,\Delta}) = \Delta + 2$. If we label the cycle $C_n$, then we get, $\lambda_2^t(C_n) = 4$.

When a graph contains two or more cycles joined with a common cutvertex, then the value of $\lambda_2^t$ equal to $\Delta + 2$, if all cycles are of even lengths and $\Delta + 1$, for others.

Let $G$ be a graph, contains a cycle of any length and finite number of edges and they have a common cutvertex $v_0$. If $\Delta$ be the degree of the cutvertex, then $\lambda_2^t(G) = \Delta + 2$, if the cycle is of even length and $\Delta + 1$, for other cases.

For any sun $S_{2n}$, the value of $\lambda_2^t$ is $\Delta + 2$. If graph is obtained from $S_{2n}$ by adding an edge to each of the pendent vertex of $S_{2n}$, then $\lambda_2^t = \Delta + 2$ for that graph. For a graph which contains a cycle of any length and each vertex of the cycle contain another cycle of any length, then $\lambda_2^t$ equal to $\Delta + 2$. The $\lambda_2^t$ value of the path, caterpillar graph and lobster are same and equal to $\Delta + 2$. One of the important result of $(2,1)$-total labelling of cactus graph is described below.

Let $G_1$ and $G_2$ be two cactus graphs. If $\Delta_1 + 1 \leq \lambda_2^t(G_1) \leq \Delta_1 + 2$ and $\Delta_2 + 1 \leq \lambda_2^t(G_2) \leq \Delta_2 + 2$, then $\Delta + 1 \leq \lambda_2^t(G) \leq \Delta + 2$, $G$ is the union of two graphs $G_1$ and $G_2$, they have only one common vertex $v$ and max $\{\Delta_1, \Delta_2\} \leq \Delta \leq \Delta_1 + \Delta_2$.

Combining all the results, we conclude that

If $\Delta$ is the degree of a cactus graph $G$, then $\Delta + 1 \leq \lambda_2^t(G) \leq \Delta + 2$.